\documentclass[aps,prb,reprint,groupedaddress]{revtex4-2}

\usepackage{amssymb}
\usepackage{amsmath}
\usepackage{graphicx}

\usepackage{txfonts}

\begin{document}

\title{Spontaneous magnetization of Kagome lattice in Ising model}
	
\author{F.A. Kassan-Ogly}
\email{Felix.Kassan-Ogly@imp.uran.ru}
\affiliation{M.N. Mikheev Institute of Metal Physics of Ural Branch of Russian Academy of Sciences, 620108, Ekaterinburg 620108, Russia}

%\date{\today}

\begin{abstract}
The spontaneous magnetization of the Kagome lattice in the Ising model is investigated. The proof of the fallacy of spontaneous magnetization obtained earlier and repeatedly migrating from publication to publication is given. An exact expression is presented in the standard Onsager form for spontaneous magnetization in the anisotropic case (depending on arbitrary values and signs of exchange interactions in all three directions in the Kagome lattice).
\end{abstract}

\maketitle

\section{Introduction}
	
The Kagome lattice was discovered by Johannes Kepler in 1619~\cite{1}, as one of the eleven Archimedean lattices. It received its name, meaning a woven bamboo basket, in 1951 in the work of Itiro Syozi~\cite{2}. It attracted the attention of researchers after Kenzi Kan\^{o} and Shigeo Naya obtained in 1953~\cite{3} an exact analytical solution for the Helmholtz free energy, namely, the principal eigenvalue of the transfer matrix by the Kramers-Wannier method, in the Ising model~\cite{4}. One of the remarkable features of the Kagome lattice is the fact that at antiferromagnetic exchange interactions, frustrations form in the system and the phase transition disappears, and this behavior is characteristic of spin systems, such as quantum spin liquids.

For many years, this circumstance has served numerous searches for the realization of this phenomenon in specific 3D materials. In 2005, Shores et al.~\cite{5} synthesized a high-purity crystal of the rare mineral herbertsmithite $\mathrm{ZnCu_3(OH)_6Cl_2}$, in which copper ions form 2D layers with an ideal Kagome lattice. Studies of magnetic properties have shown that, despite strong exchange interactions, herbertsmithite does not achieve magnetic ordering up to very low temperatures. This was followed by a rapid growth of research of herbertsmithite and related compounds in which magnetic ions occupy the Kagome lattice. To date, many articles on this topic have already been published (see, for example,~\cite{6,7} with hundreds of references therein).
The aim of this work is a refined study of spontaneous magnetization on the Kagome lattice in the Ising model.

\begin{figure}[ht]
\centering\includegraphics[width=0.8\linewidth]{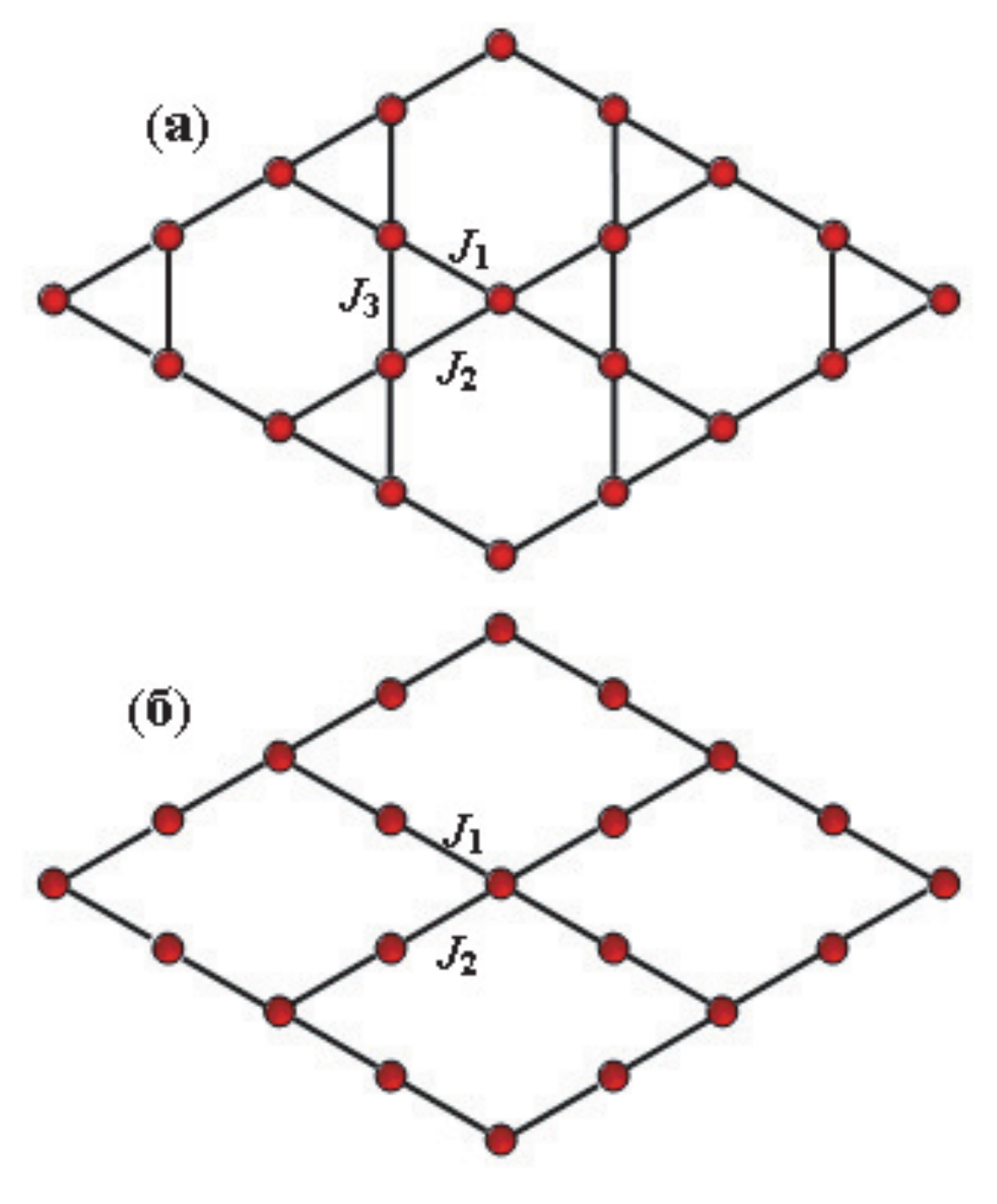}
\caption{Kagome lattice (a). Kagome lattice at $J_3=0$ (b)}
\label{fig1}
\end{figure}

\section{Spontaneous magnetization}

The outstanding work of Lars Onsager~\cite{8}, in which for the first time an exact solution for the Helmholtz free energy on a square lattice in the Ising model was obtained, actually defined the boundary of demarcation between the previous naive and very simplified theories of the mean (molecular) field type and the modern era of the study of critical phenomena in statistical physics with accompanying peculiar features of the behavior of thermodynamic quantities. The most significant consequence of the exact solution is the logarithmic (Onsager) singularity of the heat capacity near the phase transition point, observed in all subsequent exact solutions on all planar lattices without exception. The intractability of the highly complicated original Onsager method based on Lie algebra prompted subsequent researchers to search for simpler mathematical methods, and they were crowned with success. Onsager's results were re-evaluated in various combinatorial approaches developed, in particular, by Katz and Ward~\cite{9}, Potts and Ward~\cite{10}, Hurst and Green~\cite{11}, Vdovichenko~\cite{12,13}; fermionization by Schultz, Lieb and Mattis~\cite{14};  in later theories based on Grassmann variables; by Samuel~\cite{15}, Plechko~\cite{16}, and also using the Clifford-Dirac algebra by Vergeles~\cite{17}.

The beginning of the study of the spontaneous magnetization in the Ising model on two-dimensional lattices, defined as the square root of the paired spin-spin correlation function with the distance between the spins tending to infinity~\cite{18,19}, was also initiated by Onsager in 1948, when at the conference at Cornell University he simply wrote down his famous formula with chalk on the blackboard (the second outstanding Onsager's result)
\begin{equation}
M_{\text{Ons}}=\left[1-\sinh^{-2}\left(\frac{2J_{1}}{T}\right)\sinh^{-2}\left(\frac{2J_{2}}{T}\right)\right]^{1/8}.
\label{1}
\end{equation}

Later, in 1949, at the conference on Statistical Mechanics held under the auspices of the International Union for Pure and Applied Physics (IUPAP) in Florence, he announced that he and Kaufman had solved this problem. However, Onsager and Kaufman never published the derivation of this formula. Subsequently, the derivation was made by Young in 1952~\cite{20}, but for a simpler (isotropic) case and in a different form:
\begin{equation}
M_{\text{Y}}=\left[\frac{1+x^{2}}{\left(1-x^{2}\right)^{2}}\left(1-6x^{2}+x^{4}\right)^{1/2}\right]^{1/4},
\label{2}
\end{equation}
here $x=e^{\frac{-2J}{T}}$.

Yang's derivation turned out to be very complicated and, moreover, without the possibility of generalization to the anisotropic case and other lattices. The generalization to the anisotropic case was performed by Chang~\cite{21}
\begin{equation}
M_{\text{Ch}}=\left[1-\left(\frac{2x_{1}}{1-x_{1}^{2}}\right)\left(\frac{2x_{2}}{1-x_{2}^{2}}\right)\right]^{1/8},
\label{3}
\end{equation}
here $x_{1}=e^{\frac{-2J_{1}}{T}}$, $x_{2}=e^{\frac{-2J_{2}}{T}}$.

Later, Montroll, Potts and Ward once again derived Onsager's result by a much simplified method using Pfaffians and Toeplitz determinants~\cite{19}
\begin{equation}
M_{\text{MPW}}=\left[1-\frac{\left(1-z_{1}^{2}\right)^{2}\left(1-z_{2}^{2}\right)^{2}}{16z_{1}^{2}z_{2}^{2}}\right]^{1/8},
\label{4}
\end{equation}
here $z_{1}=\tanh\left(\frac{J_{1}}{T}\right)$, $z_{2}=\tanh\left(\frac{J_{2}}{T}\right)$.

By simple algebraic transformations, we were convinced that all three formulae for spontaneous magnetization of Onsager, Chang and Montroll--Potts--Ward \eqref{1}, \eqref{3} and \eqref{4} are equivalent, that is, for any values and signs of exchange interactions, there is a coincidence of the spontaneous magnetization at any temperature. In particular, the coincidence of the temperature of the spontaneous magnetization vanishing with the phase transition point calculated by the expression for the free energy from the work of Onsager~\cite{8}.

\subsection{Spontaneous magnetization on triangular and hexagonal lattices}

After obtaining, following Onsager, an exact solution for the free energy in the Ising model on a triangular lattice in 1950 by Wannier~\cite{22}, the studies of spontaneous magnetization in the Ising model were undertaken by various methods and authors, in particular, by Potts~\cite{23}, Stephenson~\cite{24}, Syozi and Naya~\cite{23}. All expressions for spontaneous magnetization obtained in these papers are presented in the same way as in the formulae for a square lattice, through hyperbolic sines and cosines, exponents and hyperbolic tangents. Similar studies were performed on a hexagonal lattice~\cite{24,25,26}. 

With the help of similar simple algebraic transformations of all the formulae obtained for spontaneous magnetization on triangular and hexagonal lattices, we were convinced of their complete equivalence, and this may indicate in favor of their verity.
However, it should be noted that they are obtained using various additional assumptions and without any rigorous mathematical justification. In other words, there are currently no algorithms for deriving spontaneous magnetization, even if there is an exact solution for free energy, as there are no criteria for the verity of expressions for spontaneous magnetization, except for two natural conditions. The first necessary condition is the obligatory coincidence of the phase transition temperature, obtained from exact solutions for free energy, with the temperature of the spontaneous magnetization vanishing. The second necessary condition is that when the temperature tends to zero, spontaneous magnetization should tend to unity.

\subsection{Spontaneous magnetization on Kagome lattice}
	
An analysis of the data available in the literature on studies of spontaneous magnetization in the Ising model showed that, in contrast to the square, triangular and hexagonal, the situation on the Kagome lattice is by no means so successful.

Immediately after obtaining an exact solution for the free energy in the Ising model on the Kagome lattice, Naya made the first attempt to study spontaneous magnetization on this lattice and obtained the first expression for the simple (isotropic) case when the exchange interactions in all three directions in the Kagome lattice are the same~\cite{26}
\begin{multline}
M_{\text{N}}=\frac{\left(1+3u^{2}\right)^{1/2}\left(1-u^{2}\right)^{1/2}}{1+u^{2}}\\
\times\left[1-\frac{128u^{6}\left(1+u^{2}\right)^{3}\left(1+3u^{4}\right)}{\left(1-u^{2}\right)^{6}\left(1+3u^{2}\right)^{2}}\right]^{1/8},
\label{5}
\end{multline}
here $u=e^{\frac{-2J}{T}}$.

In 1955, Syozi and Nakano~\cite{27} presented this formula once again, but without the multiplier in front of the square bracket.
	
The generalization of spontaneous magnetization to the anisotropic case was performed by Syozi and Naya in 1960~\cite{25},  but in a completely new form and with a new multiplier before the square bracket
\begin{equation}
M_{\text{SN}}=\frac{1}{3}\left(\sqrt{\frac{\Delta}{\Delta_{1}}}+\sqrt{\frac{\Delta}{\Delta_{2}}}+\sqrt{\frac{\Delta}{\Delta_{3}}}\right)\left[1-K_{k}\right]^{1/8},
\label{6}
\end{equation}
\begin{multline}
K_{k}=\frac{16}{\Delta^{6}}\left[\Delta^{2}(s_{1}^{2}s_{2}^{2}\Delta_{1}^{2}\Delta_{2}^{2}+s_{2}^{2}s_{3}^{2}\Delta_{2}^{2}\Delta_{3}^{2}+s_{3}^{2}s_{1}^{2}\Delta_{3}^{2}\Delta_{1}^{2}) 
\vphantom{\left(\prod\limits _{1}^{3}\right)}\right.\\
+\left.8\prod\limits _{i=1}^{3}s_{i}^{2}\Delta_{i}+\prod\limits _{i=1}^{3}s_{i}\sqrt{\Delta_{i}}(s_{i}^{2}+\Delta_{i})\right],
\label{7}
\end{multline}
\begin{equation}
\Delta=s_{1}^{2}s_{2}^{2}+s_{2}^{2}s_{3}^{2}+s_{3}^{2}s_{1}^{2}+2c_{1}c_{2}c_{3}s_{1}s_{2}s_{3}+2s_{1}^{2}s_{2}^{2}s_{3}^{3},
\label{8}
\end{equation}
\begin{equation}
\Delta_{i}=\Delta+s_{i}^{2},\quad s_{i}=\sinh\left(\frac{2J_{i}}{T}\right),\quad c_{i}=\cosh\left(\frac{2J_{i}}{T}\right).
\label{9}
\end{equation}

It should be noted that later, in 1972, Syozi~\cite{29} introduced some amendments to the formula~\eqref{7}, namely, made a replacement
$\Delta_{i}^{2}$ by $\Delta_{i}$. We also should note Lin's article~\cite{30}, in which he gives an expression for spontaneous magnetization that actually coincides with what is presented by Syozi and Naya~\cite{25}.

We give another expression for spontaneous magnetization for the isotropic case (formula 4.15) from the article by Matveev and Shrock~\cite{31}.
\begin{equation}
M_{\text{MS}}=\frac{\left(1-6u-3u^{2}\right)^{1/8}\left(1+2u+5u^{2}\right)^{3/8}\left(1+3u^{2}\right)^{1/4}}{(1-u)^{1/4}(1+u)},
\label{10}
\end{equation}
here $u=e^{\frac{-4J}{T}}$.

\begin{figure}[ht]
\centering\includegraphics[width=0.8\linewidth]{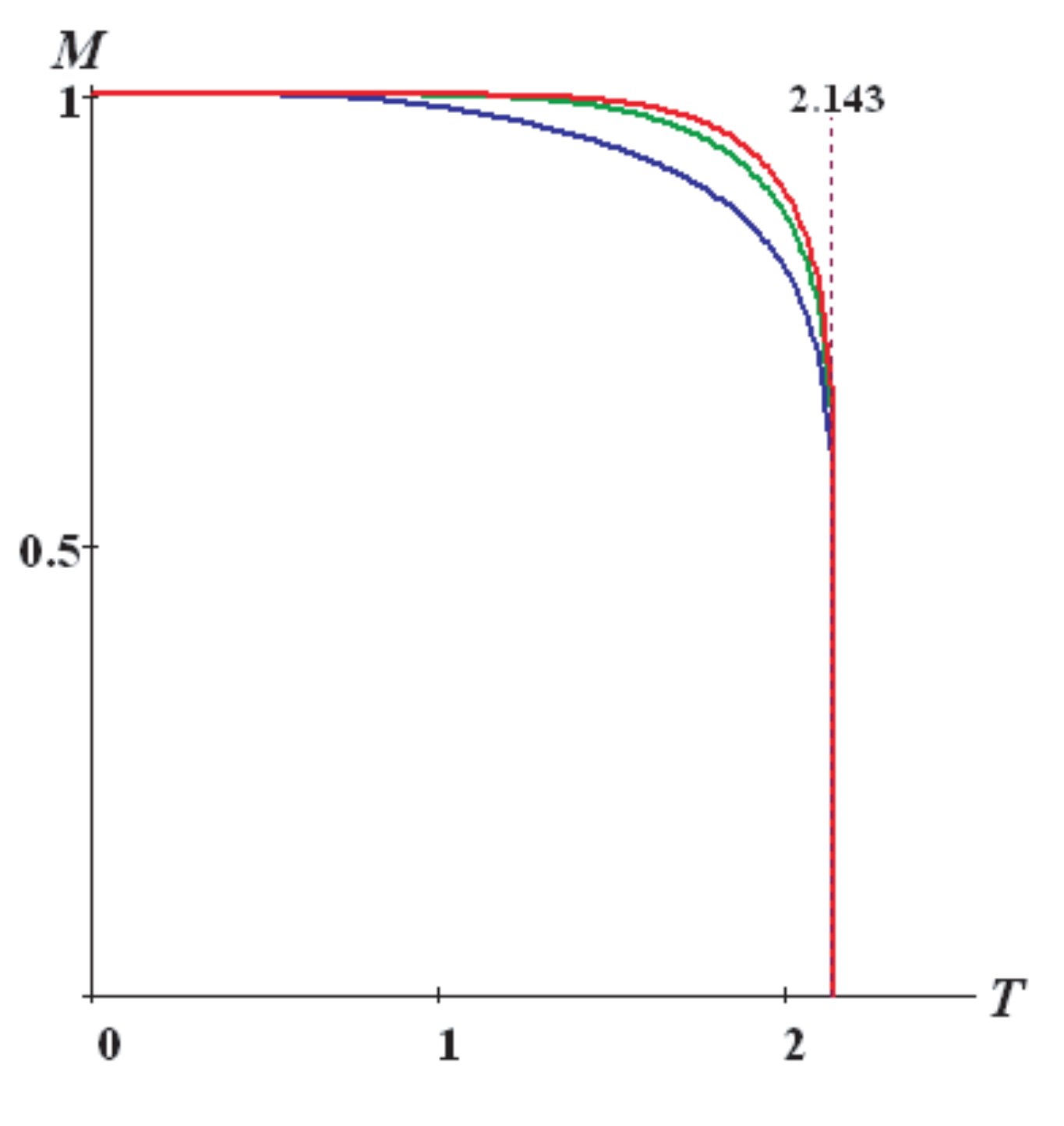}
\caption{Spontaneous magnetization on Kagome lattice at $J_{1}=J_{2}=J_{3}=1$}
\label{fig2}
\end{figure}

Let us analyze a special (isotropic) case when all the exchange interactions are ferromagnetic and the same: $J_{1}=J_{2}=J_{3}=1$. The results of numerical calculations for all the above formulae showed the following. Firstly, all spontaneous magnetizations are \textit{different}. Secondly, in the formula belonging to Syozi and Naya~\cite{28}, and repeated by Lin~\cite{30},  the expression $1-K_{k}$ under the sign of the root of the eighth degree is negative at any temperature, and, as a consequence, spontaneous magnetization is purely imaginary, that is, in fact, the formula~\eqref{6} is not physical, or, simply, erroneous. Figure~\ref{fig2} shows three curves for spontaneous magnetization calculated from the results of the work by Syozi and Nakano~\cite{27} (red curve), Syozi~\cite{29} (green curve), and Matveev and Shrock~\cite{31} (blue curve). Note that both the first and second necessary conditions discussed earlier are met on all three curves. When the temperature tends to zero, spontaneous magnetization reaches saturation, and the temperature of the reversal of spontaneous magnetization coincides with the phase transition point defined in the article by Kan\^{o} and Naya~\cite{3} $T_{\text{c}}=4/\ln(3+2\sqrt{3})$.

Nevertheless, in all intermediate temperatures, the spontaneous magnetizations on the three curves do not coincide. Thus, the problem arises which formula for spontaneous magnetization of the three presented is true, or even all of them are erroneous.

\begin{figure}[ht]
\centering\includegraphics[width=0.8\linewidth]{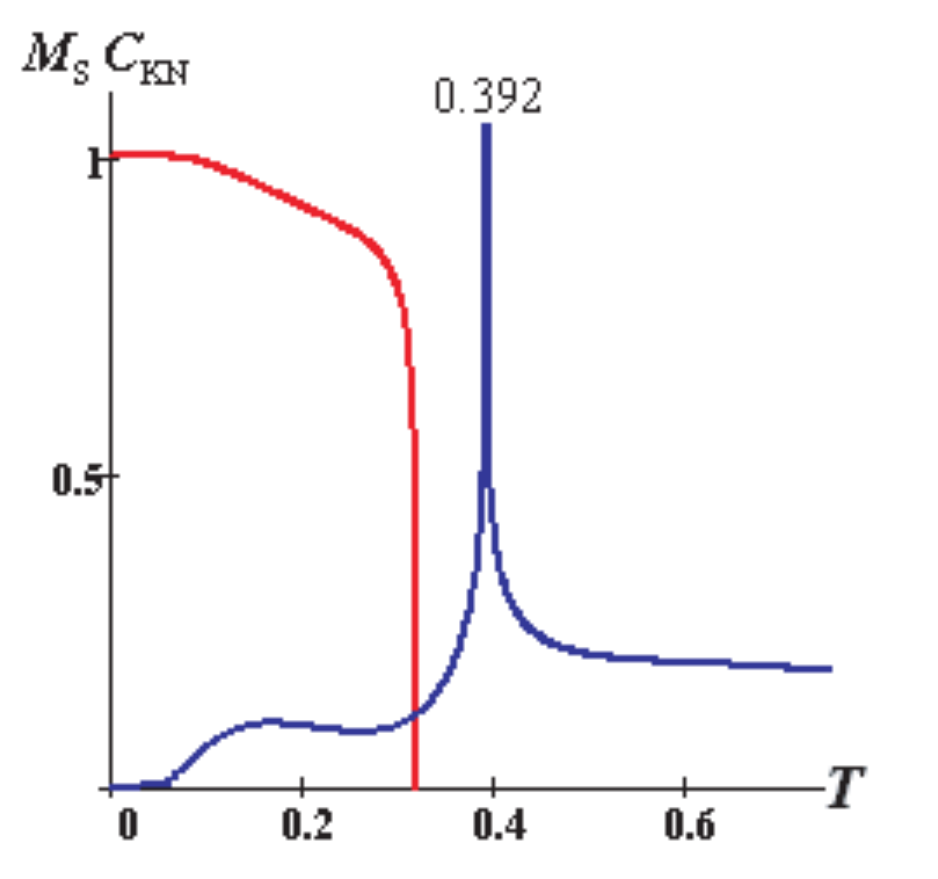}
\caption{Spontaneous magnetization and heat capacity on Kagome lattice at $J_{1}=0.2$, $J_{2}=0.3$, $J_{3}=-0.8$}
\label{fig3}
\end{figure}

Let us now consider additionally the general (anisotropic) case when the exchange interactions $J_{1}$, $J_{2}$, and $J_{3}$, differ both in signs and in magnitude. Firstly, multiple numerical calculations using the formula belonging to Syozi and Naya~\cite{28} and Lin~\cite{30} show that for all signs and quantities and at all temperatures, spontaneous magnetization is purely imaginary, that is, in fact, this formula is absolutely erroneous. Secondly, we present the calculation of spontaneous magnetization according to the corrected Syozi~\cite{29} formula for the following exchange interactions: $J_{1}=0.2$, $J_{2}=0.3$, and $J_{3}=-0.8$. In the Figure~\ref{fig3}, the red curve indicates the spontaneous magnetization calculated for such interactions, and the blue curve indicates the heat capacity calculated according to standard statistical physics formulae using free energy from the article by Kan\^{o} and Naya~\cite{3}. In this case, the phase transition temperature is $\approx0.392$. From the Figure~\ref{fig3} it is clearly seen that the point of spontaneous magnetization vanishing does not coincide with the phase transition temperature, that is, the first necessary condition is violated. And this is already sufficient evidence of the fallacy of the Syozi's formula~\cite{29}.

\begin{figure}[ht]
\centering\includegraphics[width=0.8\linewidth]{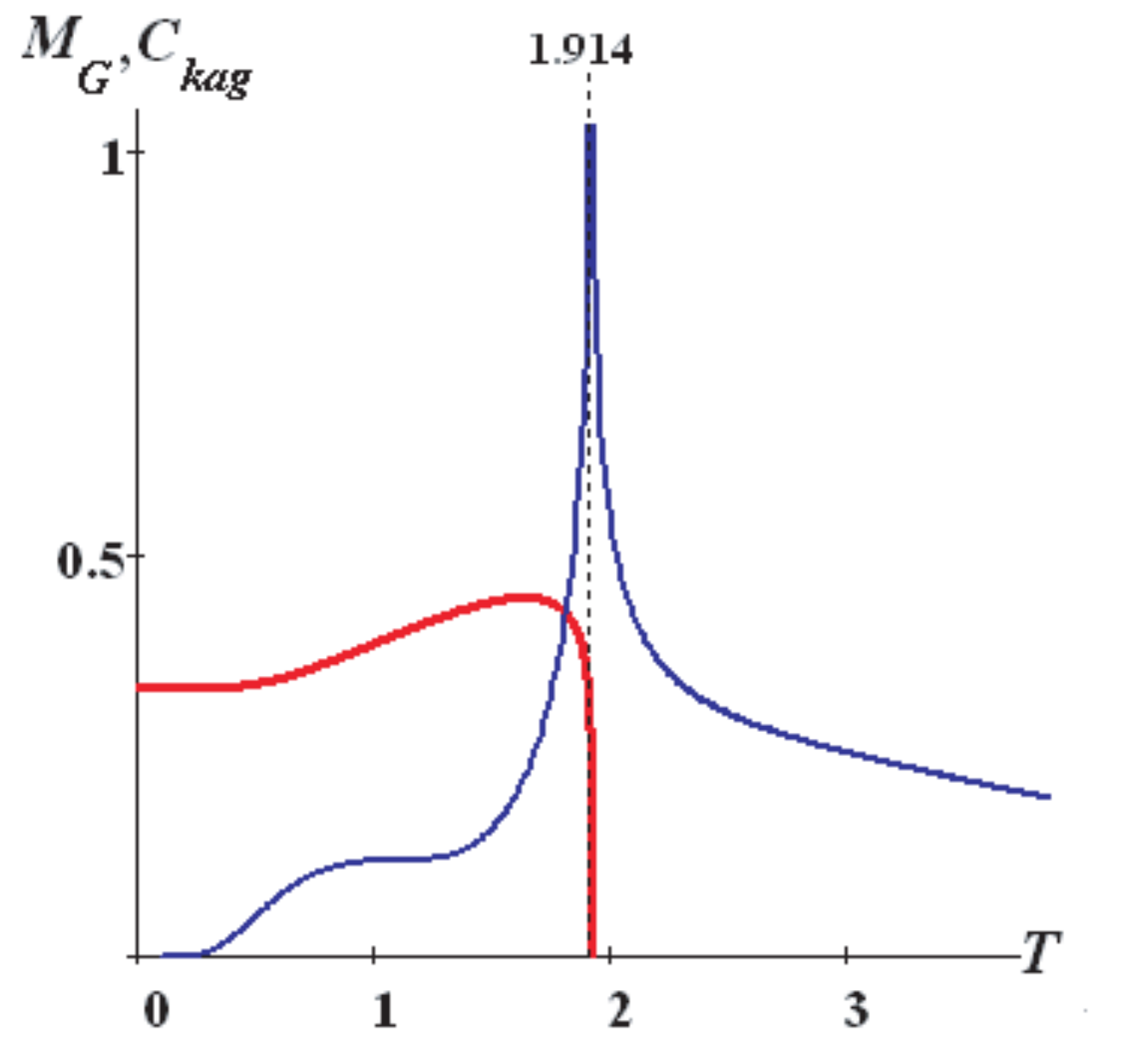}
\caption{Spontaneous magnetization and heat capacity on Kagome lattice at $J_{1}=0.3$, $J_{2}=0.3$, $J_{3}=2.9$}
\label{fig4}
\end{figure}

To complete the picture, we present the calculation of spontaneous magnetization according to Ge\u{\i}likman's~\cite{32} formulae. In the Figure~\ref{fig4}, the red curve indicates the spontaneous magnetization calculated for exchange interactions: $J_{1}=0.3$, $J_{2}=0.3$, and $J_{3}=2.9$, and the blue curve indicates the heat capacity calculated according to standard formulae of statistical physics using free energy from the article by Kan\^{o} and Naya~\cite{3}. From the Figure~\ref{fig4} it can be seen that the point of spontaneous magnetization vanishing coincides with the temperature of the phase transition, that is, the first necessary condition is not violated, but the second condition is violated~--- spontaneous magnetization does not reach saturation when the temperature tends to zero. This is sufficient evidence of the fallacy of the Ge\u{\i}likman's formula~\cite{32}.

We suggested the corrected expression for spontaneous magnetization in the Ising model on Kagome lattice: 
\begin{equation}
M_{\text{kag}}=\left[1-K_{k}\right]^{1/8},
\label{11}
\end{equation}
\begin{multline}
K_{k}=\frac{16}{K^{6}}\left[K^{2}(s_{1}^{2}s_{2}^{2}K_{1}^{2}K_{2}^{2}+s_{2}^{2}s_{3}^{2}K_{2}^{2}K_{3}^{2}+s_{3}^{2}s_{1}^{2}K_{3}^{2}K_{1}^{2}) 
\vphantom{\prod\limits _{i=1}^{3}}\right.\\
+\left.8\prod\limits _{i=1}^{3}s_{i}^{2}K_{i}^{2}-\prod\limits _{i=1}^{3}s_{i}K_{i}(s_{i}^{2}+K_{i}^{2})\right],
\label{12}
\end{multline}
\begin{multline}
K=\frac{1}{16u_{1}^{2}u_{2}^{2}u_{3}^{2}}(u_{1}+u_{2}+u_{3}+u_{1}u_{2}u_{3})(u_{1}-u_{2}-u_{3}+u_{1}u_{2}u_{3})\\
\times(u_{2}-u_{3}-u_{1}+u_{1}u_{2}u_{3})(u_{3}-u_{1}-u_{2}+u_{1}u_{2}u_{3}),
\label{13}
\end{multline}
\begin{equation}
K_{1}=\frac{-u_{1}^{2}+u_{2}^{2}+u_{3}^{2}-u_{1}^{2}u_{2}^{2}u_{3}^{2}}{4u_{1}u_{2}u_{3}},
\label{14}
\end{equation}
\begin{equation}
K_{2}=\frac{u_{1}^{2}-u_{2}^{2}+u_{3}^{2}-u_{1}^{2}u_{2}^{2}u_{3}^{2}}{4u_{1}u_{2}u_{3}},
\label{15}
\end{equation}
\begin{equation}
K_{3}=\frac{u_{1}^{2}+u_{2}^{2}-u_{3}^{2}-u_{1}^{2}u_{2}^{2}u_{3}^{2}}{4u_{1}u_{2}u_{3}},
\label{16}
\end{equation}
\begin{equation}
u_{i}=e^{2J_{i}/T},\quad s_{i}=\sinh\left(\frac{2J_{i}}{T}\right).
\label{17}
\end{equation}

Let us give a proof of the validity of formulae \eqref{11}--\eqref{17}.  To do this, consider again the Kagome lattice, but in which the exchange interaction $J_3$ in one of the directions is zero (see Figure~\ref{fig1}). It clearly follows from this drawing that in such an extreme case, an undecorated Kagome lattice exactly coincides with a square lattice singly decorated in both two directions. Moreover, the interactions between nodal spins in such a square lattice are zero.

\begin{figure}[ht]
\centering\includegraphics[width=0.8\linewidth]{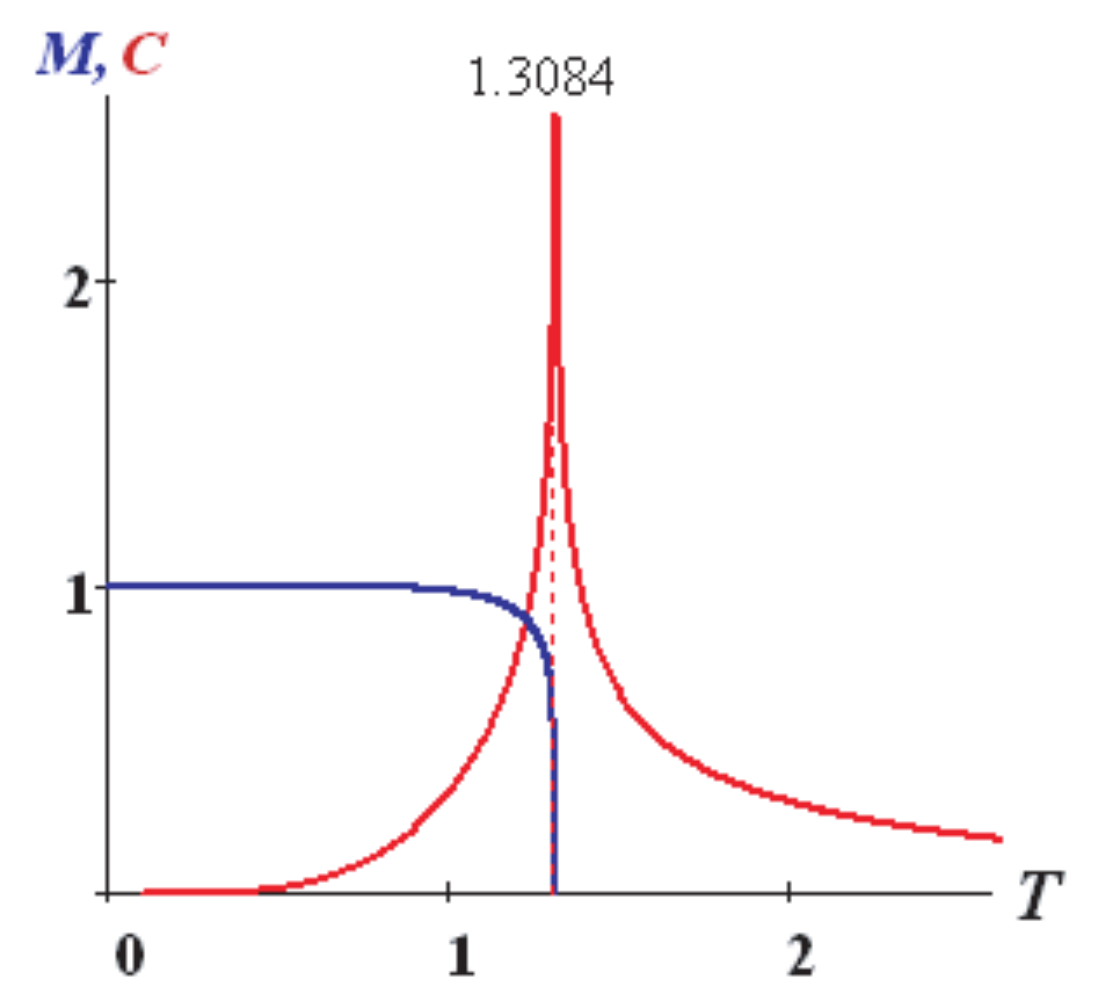}
\caption{Spontaneous magnetization and heat capacity on decorated square lattice at $J_{1d}=J_{2d}=1$ and on undecorated Kagome lattice at $J_{1}=J_{2}$, $J_{3}=0$}
\label{fig5}
\end{figure}

We performed multiple calculations of the heat capacity using two formulae: for the free energy of the Kagome lattice (from the article by Kan\^{o} and Naya~\cite{3}) with interactions only between nodal spins $J_{1}$, $J_{2}$ and $J_{3}=0$ and for the square lattice~\cite{33,34} with interactions only between decorating spins $J_{1d}$, $J_{2d}$ (without interactions between nodal spins) for a variety of values and signs of all interactions: $J_{1}$, $J_{2}$, $J_{1d}$ and $J_{2d}$ and made sure that in all cases the results are the same. As an example, Figure~\ref{fig5} shows calculation curves for a special case $J_{1}=J_{2}=J_{1d}=J_{2d}=1$.

The curves of these two calculations are simply indistinguishable. It follows from this that the calculations of spontaneous magnetization (if it is correct) in these lattices should coincide, which is confirmed by direct calculations using our formulae \eqref{11}--\eqref{17} for the undecorated Kagome lattice and the formulae for the decorated square lattice
\begin{equation}
M_{\text{sq}}=\left[1-\frac{D_{1}^{2}D_{2}^{2}}{S_{1}^{2}S_{2}^{2}}\right]^{1/8},
\label{18}
\end{equation}
\begin{equation}
D_{i}=\cosh^{2d_{i}+2}\left(\frac{2J_{id}}{T}\right)-\sinh^{2d_{i}+2}\left(\frac{2J_{id}}{T}\right),
\label{19}
\end{equation}
\begin{multline}
S_{i}=\frac{e^{2J_{id}/T}}{2}\left[\cosh^{d_{i}+1}\left(\frac{J_{id}}{T}\right)+\sinh^{d_{i}+1}\left(\frac{J_{id}}{T}\right)\right]^{2}\\
-\frac{e^{-2J_{id}/T}}{2}\left[\cosh^{d_{i}+1}\left(\frac{J_{id}}{T}\right)-\sinh^{d_{i}+1}\left(\frac{J_{id}}{T}\right)\right]^{2},
\label{20}
\end{multline}
here $d_i$ is the decoration degree, and in our case $d_{1}=d_{2}=1$, $J_{1d}=J_{2d}=1$ and $J_{1}=J_{2}=0$.

The curves of these two calculations of spontaneous magnetization, shown in Figure~\ref{fig5}, are also indistinguishable. The phase transition point, determined from the heat capacity, is equal to $2/\ln(\sqrt{2\sqrt{2}+2}+\sqrt{2}+1)\approx1.308$, and exactly coincides with the point of vanishing of spontaneous magnetization, which corresponds to the fulfillment of the first necessary condition discussed above. It follows from Figure~\ref{fig5} that, naturally, the second necessary condition is fulfilled~--- spontaneous magnetization reaches saturation when the temperature tends to zero.

Summarizing the results of our calculations, we conclude that the fulfillment of the necessary conditions and the coincidence of the spontaneous magnetization of the decorated square lattice and the undecorated Kagome lattice at all temperatures is a proof of the verity of our formulae \eqref{11}--\eqref{17}.
Multiple calculations of spontaneous magnetization according to our formulae for various values and signs of exchange interactions have shown that the necessary conditions are never violated.

\subsection{Frustration properties in Ising model on Kagome lattice}

Consider one remarkable phenomenon observed in the Ising model on the Kagome lattice. At some values and signs of exchange interactions $J_{1}$, $J_{2}$ and $J_{3}$, strong frustrations arise in it, leading to degeneracy of the ground state, which is accompanied by a nonzero entropy value in the ground state (at~$T=0$). These frustrations lead to a complete disappearance of spontaneous magnetization. When using our formulae \eqref{11}--\eqref{17}, it is found that frustrations occur under the following conditions. Firstly, all three interactions must necessarily be equal in modulus. Secondly, either all interactions must be antiferromagnetic, or one of them antiferromagnetic and the rest two ferromagnetic. At any violation of these conditions, frustrations disappear, a phase transition and spontaneous magnetization occur in full accordance with our formulae \eqref{11}--\eqref{17} and with the exact solution for free energy obtained in the article by Kan\^{o} and Naya~\cite{3}.
It is noteworthy that for any values of the parameters leading to frustrations, the entropy in the ground state (residual entropy) has the same universal value, namely,
\begin{multline}
S_{0}=\frac{1}{6\pi}\int\limits _{0}^{\pi}\ln\left[\frac{21+5\sqrt{17}}{4}
\vphantom{\sqrt{\left(\frac{8}{8\cos^{2}\alpha}\right)^{2}}} \right.\\
\times\left.\left[1+\sqrt{1-\left(\frac{8\cos\alpha}{25-8\cos^{2}\alpha}\right)^{2}}\right]\right]d\alpha\approx0.501\,183\,3.
\label{21}
\end{multline}
The study of the phenomenon of frustrations when using the exact solution for free energy in the Ising model on the Kagome lattice, obtained by Kan\^{o} and Naya~\cite{3}, leads to the same results, which also testifies to the correctness of our formulae.

Using our formulae \eqref{11}--\eqref{17}, we found a novel type of frustration in the Ising model on the Kagome lattice, which occurs under the following conditions. Namely, if we choose one of the exchange interactions ferromagnetic $J_{1}>0$, the second also ferromagnetic $J_{2}>0$, the third antiferromagnetic $J_{3}<0$, and the second and third are the same modulo, but less than the first modulo, then for any relationship $J_{2}$ and $J_{1}$ the entropy in the ground state (residual entropy) will take the same universal value, namely, $\ln2/3$.

\begin{figure}[ht]
\centering\includegraphics[width=0.8\linewidth]{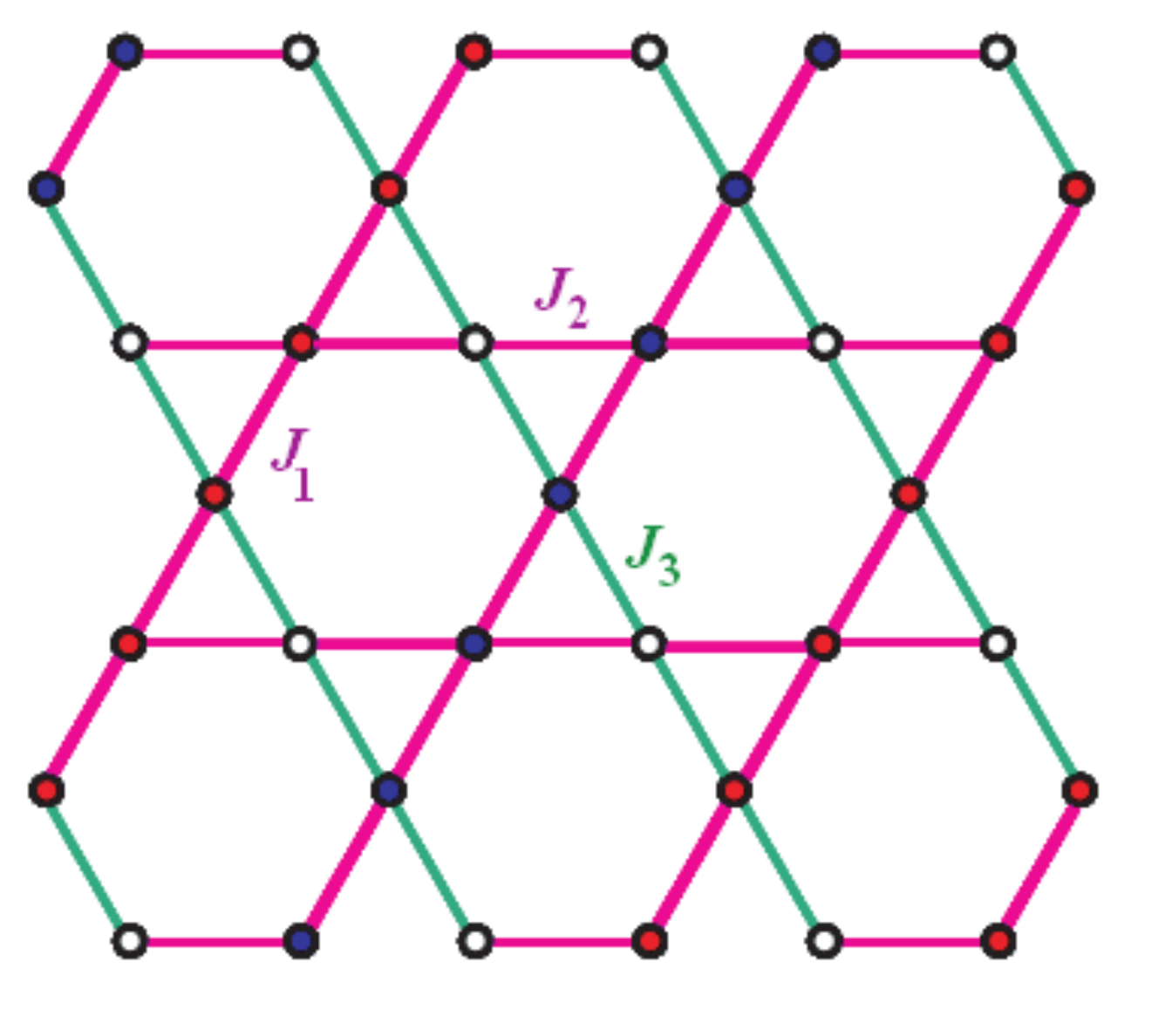}
\caption{Magnetic structure in the ground state on Kagome lattice at $J_{1}>0$, $J_{2}>0$, $J_{3}=-J_{2}$, $J_{1}>J_{2}$}
\label{fig6}
\end{figure}

Figure~\ref{fig6} shows one of the many spin configurations (magnetic structure) in the ground state. Ferromagnetic interactions are depicted in purple, antiferromagnetic~--- green. The spins pointing upwards are depicted with a red circle, the spins pointing downwards with a blue circle; with light circles~--- spins in an uncertain state (flipping spins).

\begin{figure}[ht]
\centering\includegraphics[width=0.8\linewidth]{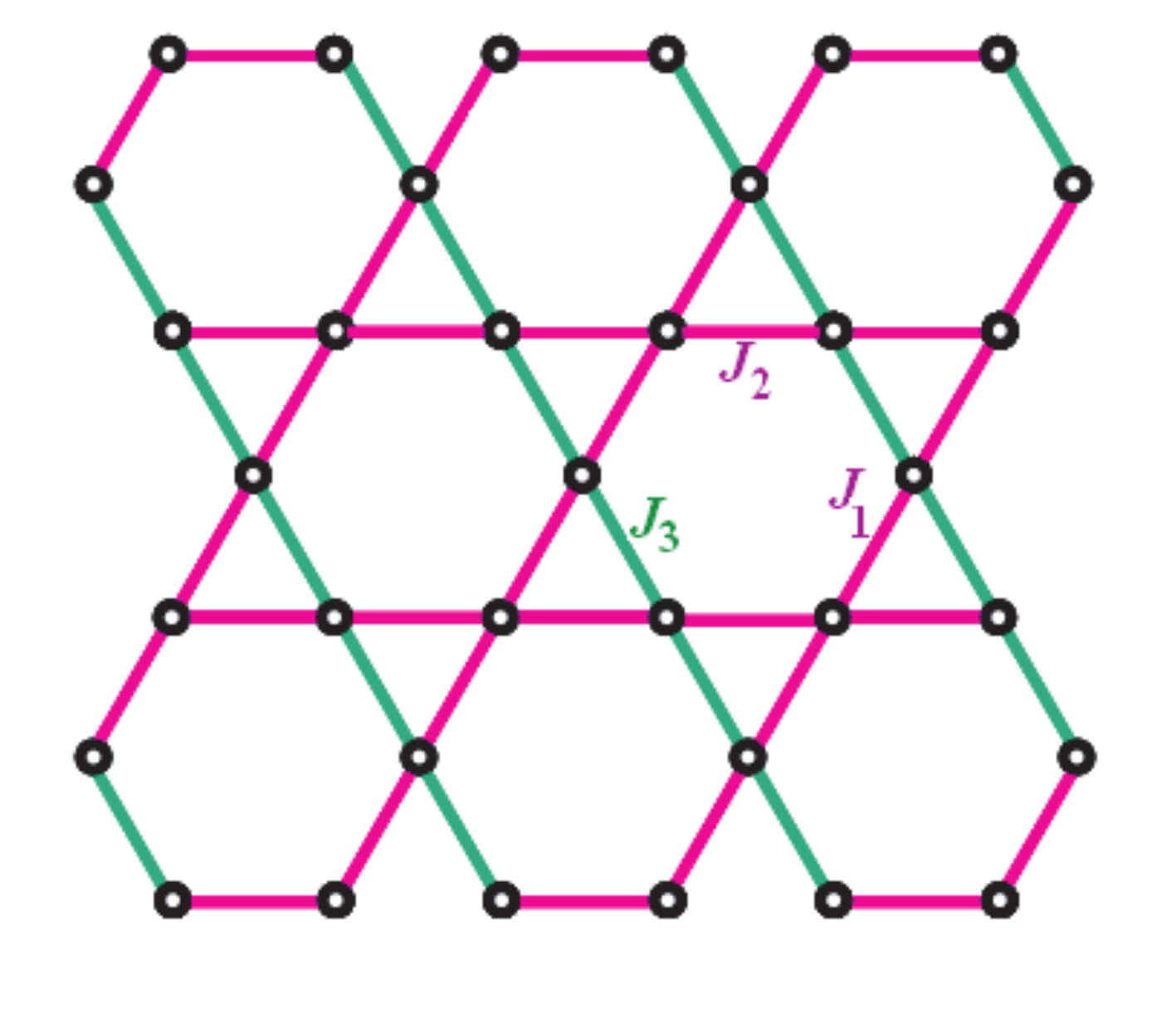}
\caption{Magnetic structure in the ground state on Kagome lattice at $J_{1}>0$, $J_{2}>0$, $J_{3}=-J_{2}$, $J_{1}=J_{2}$}
\label{fig7}
\end{figure}

From the Figure~\ref{fig6} it follows that along each chain associated with a strong ferromagnetic interaction $J_{1}>0$ all spins are directed either up or down, that is, they are ferromagnetically ordered; in other words, the translational invariance for one lattice period in one of the directions is observed. There is no ordering along the chains associated with weak interactions. In other words, this type of frustration is accompanied by the type of ordering called partial order with dimensionality reduction (see, for example,~\cite{35,36}). In the particular case, when all three exchange interactions are equal modulo, the whole picture changes abruptly, so that the ordering disappears, since all the lattice spins are in an uncertain state (Figure~\ref{fig7}). The residual entropy also changes abruptly, regaining the expression \eqref{21}.

\section{Conclusions}

The formulae of spontaneous magnetization derived in this article in the Ising model on Kagome lattice for arbitrary values of exchange interactions allow us to draw the following conclusions. Firstly, all the formulae of spontaneous magnetization on Kagome lattice available in the literature are erroneous. Secondly, in some works, even the necessary condition that spontaneous magnetization must obey is violated~--- the coincidence of the phase transition temperature obtained from exact solutions for free energy with the temperature of the spontaneous magnetization vanishing. Thirdly, despite the absence of general verity criteria, the proof of the validity of our formula is given, which allows us to hope that in any other specific case it is possible and necessary to find such a criterion.

\begin{acknowledgments}
The research was carried out within the state assignment of Ministry of Science and Higher Education of the Russian Federation (theme “Quantum” No. 122021000038-7).
\end{acknowledgments}

% BibTeX
\bibliographystyle{apsrev4-2}
\bibliography{K2022}

\end{document}